\documentclass[12pt]{article}
\usepackage{amsfonts}
\begin{document}
\def\a{\alpha}\def\b{\beta}\def\g{\gamma}\def\d{\delta}\def\e{\epsilon }
\def\k{\kappa}\def\l{\lambda}\def\L{\Lambda}\def\s{\sigma}\def\S{\Sigma}
\def\Th{\Theta}\def\th{\theta}\def\om{\omega}\def\Om{\Omega}\def\G{\Gamma}
\def\y{\vartheta}\def\m{\mu}\def\n{\nu}
\def\ws{worldsheet}
\def\susy{supersymmetry}
\def\ts{target superspace}
\def\ks{$\k$--symmetry}
\newcommand{\plabel}{\label}
\def\PRL #1 #2 #3{{\em Phys. Rev. Lett. \/} {\bf#1} (#2) #3}
\def\NPB #1 #2 #3{{\em Nucl. Phys. \/} {\bf B#1} (#2) #3}
\def\NPBFS #1 #2 #3 #4{{\em Nucl. Phys. \/} {\bf B#2} [FS#1] (#3) #4}
\def\CMP #1 #2 #3{{\em Commun. Math. Phys. \/} {\bf #1} (#2) #3}
\def\prd #1 #2 #3{{\em Phys. Rev. \/} {\bf D#1} (#2) #3}
\def\PLA #1 #2 #3{{\em Phys. Lett. \/} {\bf #1A} (#2) #3}
\def\plb #1 #2 #3{{\em Phys. Lett. \/} {\bf B#1} (#2) #3}
\def\JMP #1 #2 #3{{\em J. Math. Phys. \/} {\bf #1} (#2) #3}
\def\PTP #1 #2 #3{{\em Prog. Theor. Phys. \/} {\bf #1} (#2) #3}
\def\SPTP #1 #2 #3{{\em Suppl. Prog. Theor. Phys. \/} {\bf #1} (#2) #3}
\def\ap #1 #2 #3{{\em Ann. of Phys. \/} {\bf #1} (#2) #3}
\def\PNAS #1 #2 #3{{\em Proc. Natl. Acad. Sci. USA} {\bf #1} (#2) #3}
\def\RMP #1 #2 #3{{\em Rev. Mod. Phys. \/} {\bf #1} (#2) #3}
\def\PR #1 #2 #3{{\em Phys. Reports \/} {\bf #1} (#2) #3}
\def\AoM #1 #2 #3{{\em Ann. of Math. \/} {\bf #1} (#2) #3}
\def\UMN #1 #2 #3{{\em Usp. Mat. Nauk \/} {\bf #1} (#2) #3}
\def\FAP #1 #2 #3{{\em Funkt. Anal. Prilozheniya \/} {\bf #1} (#2) #3}
\def\FAaIA #1 #2 #3{{\em Functional Analysis and Its Application \/} {\bf
#1} (#2) #3}
\def\BAMS #1 #2 #3{{\em Bull. Am. Math. Soc. \/} {\bf #1} (#2)
#3} \def\TAMS #1 #2 #3{{\em Trans. Am. Math. Soc. \/} {\bf #1}
(#2) #3}
\def\InvM #1 #2 #3{{\em Invent. Math. \/} {\bf #1} (#2) #3}
\def\LMP #1 #2 #3{{\em Letters in Math. Phys. \/} {\bf #1} (#2) #3}
\def\ijmpa #1 #2 #3{{\em Int. J. Mod. Phys. \/} {\bf A#1} (#2) #3}
\def\ijmpd #1 #2 #3{{\em Int. J. Mod. Phys. \/} {\bf D#1} (#2) #3}
\def\AdM #1 #2 #3{{\em Advances in Math. \/} {\bf #1} (#2) #3}
\def\RMaP #1 #2 #3{{\em Reports on Math. Phys. \/} {\bf #1} (#2) #3}
\def\IJM #1 #2 #3{{\em Ill. J. Math. \/} {\bf #1} (#2) #3}
\def\APP #1 #2 #3{{\em Acta Phys. Polon. \/} {\bf #1} (#2) #3}
\def\TMP #1 #2 #3{{\em Theor. Mat. Phys. \/} {\bf #1} (#2) #3}
\def\JPA #1 #2 #3{{\em J. Physics \/} {\bf A#1} (#2) #3}
\def\JSM #1 #2 #3{{\em J. Soviet Math. \/} {\bf #1} (#2) #3}
\def\mpla #1 #2 #3{{\em Mod. Phys. Lett. \/} {\bf A#1} (#2) #3}
\def\JETP #1 #2 #3{{\em Sov. Phys. JETP \/} {\bf #1} (#2) #3}
\def\JETPL #1 #2 #3{{\em  Sov. Phys. JETP Lett. \/} {\bf #1} (#2) #3}
\def\PHSA #1 #2 #3{{\em Physica} {\bf A#1} (#2) #3}
\def\CQG #1 #2 #3{{\em Class. Quantum Grav. \/} {\bf #1} (#2) #3}
\def\SJNP #1 #2 #3{{\em Sov. J. Nucl. Phys. (Yadern.Fiz.) \/} {\bf #1} (#2) #3}
\def\hepth #1 {{hep-th/}{#1}}
\def\jhep #1 #2 #3{{\em JHEP \/} {\bf #1} (#2) #3}
\def\npps #1 #2 #3{{\em Nucl. Phys. Proc. Suppl. \/} {\bf #1} (#2) #3}

\begin{flushright}
Preprint DFPD 03/TH/45
\end{flushright}

\newcommand{\nn}{\nonumber\\}\newcommand{\p}[1]{(\ref{#1})}
\renewcommand{\thefootnote}{\fnsymbol{footnote}}
\begin{center}
{\Large GL Flatness of $OSp(1|2n)$ and Higher Spin Field Theory
from Dynamics in Tensorial Spaces} \footnote{Contribution to the
Proceedings of the International Workshop ``Supersymmetries and
Quantum Symmetries'' (SQS'03, Dubna, 24--29 July, 2003). }

\date{}
~\\
Mikhail Plyushchay ${}^{1,2}$, Dmitri Sorokin${}^{3,4}$ and Mirian Tsulaia ${}^{4,5,6}$\\
~\\
${}^1${\it Depart. de F\'{\i}sica, Universidad de Santiago de
Chile,
Casilla 307, Santiago 2, Chile} \\
${}^2${\it Institute for High Energy Physics,
Protvino, Russia} \\
{${}^3$ \it Institute for Theoretical Physics,
 NSC KIPT,
 Kharkov, 61108, Ukraine\\
${}^4$ Universit\'a degli Studi di Padova,
Dipartimento di Fisica ``Galileo Galilei"\\
and INFN, Sezione di Padova, Via F. Marzolo, 8, 35131 Padova,
Italia
\\
${}^5$ Bogoliubov Laboratory of Theoretical Physics, JINR,
 141980 Dubna, Russia \\
${}^6$ Institute of  Physics, GAS, 380077 Tbilisi, Georgia
}%

\end{center}

\begin{abstract}
A main purpose of this paper is to explain how the theory of
higher spin fields in flat $D=4$ space and in $AdS_4$ emerges as a
result of the quantization of a superparticle propagating in so
called tensorial superspaces which have the property of a
`generalized conformal' or simply General Linear (GL) flatness.

\end{abstract}
\thispagestyle{empty}
\section{Introduction}
We present some results of the development of particle dynamics
and field theory in tensorial spaces and explain their relation to
Higher Spin Field Theory. The plan of the paper is as follows. We
shall first provide a motivation why it is interesting to consider
the dynamics of particles, strings, etc. and field theory in
tensorial superspaces. In Section 2 we shall give the definition
of tensorial supermanifolds and discuss their implication to the
theory of higher spin fields. In Section 3 we shall introduce the
notion of $GL(2n)$ flatness of manifolds and give examples of
tensorial superspaces which possess this property, actually, the
only known non--trivial example being $OSp(1|2n)$ supergroup
manifolds. We shall then consider the dynamics of a particle in
flat tensorial space and on $Sp(4)$, and analyze constraints and
physical degrees of freedom of this object. Finally, in Sections 4
and 5 we shall quantize this system using GL flatness, find the
general solutions of the field equations of the quantum system and
demonstrate that its quantum spectrum consists of an infinite
tower of massless integer and half--integer higher spin states
which obey so called unfolded higher spin field equations in flat
$D=4$ space and/or in $AdS_4$ in the formulation of M. Vasiliev
\cite{m1}.

\section{Tensorial (super)spaces and higher spins}
We call a space tensorial if its points are parametrized by
symmetric $2n\times 2n$ matrix coordinates $x^{\a\b}=x^{\b\a}$
$(\a,\b=1,\cdots,2n)$ linearly transformed by the group $Sp(2n)$
\footnote{Generally speaking the group of linear transformations
of $x^{\a\b}$ is $GL(2n)$, but in what follows we shall always
restrict it to its subgroup $Sp(2n)$ to keep the simplectic and
related spinorial structure manifest.}.

The bosonic tensorial space can be extended to a tensorial
superspace by adding Grassmann odd directions parametrized by $N$
fermionic spinor coordinates $\theta^\a_i$ $(i=1,\cdots,N)$. Then
a group of linear transformations of $(x^{\a\b},\theta^\a_i)$
becomes $Sp(2n)\times O(N)$.

Examples of tensorial superspaces are
\begin{itemize}
\item
$OSp(N|2n)$ supergroup manifolds and, in particular, $OSp(1|4)$ to
be considered in detail below,
\item
flat tensorial superspaces invariant under centrally extended
super Poincare translations, e.g. under the $N=1$, $D=4$ super
Poincare group transformations
\begin{equation}\plabel{spt}
\d\theta^\a=\epsilon^\a,\quad \d x^{\a\b}={i\over 2}
(\theta^\a\epsilon^\b+\theta^\b\epsilon^\a)=i\theta^{\{\a}\epsilon^{\b\}}\,,\quad
\a,\b=1,2,3,4\,.
\end{equation}
The algebra of these transformations contains in addition to $D=4$
translations $P_m$ $(m=0,1,2,3)$ also six tensorial charge
generators $Z_{mn}=-Z_{nm}$
\begin{equation}\plabel{sp}
\{Q_\a,Q_\b\}=4P_{\a\b}=2P_m\gamma^m_{\a\b}+Z_{mn}\gamma^{mn}_{\a\b},
\end{equation}
where in the r.h.s. we have made the decomposition of the momentum
$P_{\a\b}=P_{\b\a}$ conjugate to $x^{\a\b}$ in a basis of $D=4$
$\gamma$--matrices.
\end{itemize}
Probably, the first who suggested a physical application of
tensorial spaces was C.~Fronsdal.

\subsection{Fronsdal's proposal of '85 -- alternative to Kaluza \& Klein}
 In his Essay of 1985 \cite{fronsdal1} Fronsdal conjectured that four--dimensional higher
spin field theory can be realized as a field theory on a
ten--dimensional tensorial manifold parametrized by the
coordinates
\begin{equation}\plabel{x}
x^{\a\b}={1\over 2}x^m\gamma_m^{\a\b}+{1\over 4}
y^{mn}\gamma_{mn}^{\a\b}, \quad m,n=0,1,2,3\,; \quad
\a,\b=1,2,3,4\,,
\end{equation}
where $x^m$ are associated with four coordinates of the
conventional $D=4$ space--time and $y^{mn}=-y^{mn}$ describe
spinning degrees of freedom.

The assumption was that by analogy with, for example, D=10 or D=11
supergravities, which are relatively simple theories but whose
dimensional reduction to four dimensions produces very complicated
extended supergravities, there may exist a theory in
ten--dimensional tensorial space whose alternative Kaluza--Klein
reduction may lead in $D=4$ to an infinite tower of fields with
increasing spins instead of the infinite tower of Kaluza--Klein
particles of increasing mass. The assertion was based on the
argument that the symmetry group of the theory should be
$OSp(1|8)\supset SU(2,2)$, which contains the $D=4$ conformal
group as a subgroup such that an irreducible (oscillator)
representation of $OSp(1|8)$ contains each and every massless
higher spin representation of $SU(2,2)$ only once. So the idea was
that using a single representation of $OSp(1|8)$ in the
ten-dimensional tensorial space one could describe  an infinite
tower of higher spin fields in $D=4$ space--time in a simpler way.
Fronsdal regarded the tensorial space as a space transforming
homogeneously under the transformations of $Sp(8)$. Ten is the
minimal dimension of such a space which can contain D=4
space--time as a subspace. For some reason Fronsdal gave only a
general definition and did not identify this ten--dimensional
space with any conventional manifolds, like the ones mentioned
above.

In his Essay Fronsdal also stressed the importance of $OSp(1|2n)$
supergroups for the description of theories with superconformal
symmetry. In the same period and later on different people studied
$OSp(1|2n)$ supergroups in various physical contexts. For
instance, $OSp(1|32)$ and $OSp(1|64)$ have been assumed to be
underlying superconformal symmetries of string- and M-theory.

\subsection{Particle dynamics in tensorial superspace}

Without relation to Higher Spins, in 1998 I. Bandos and J.
Lukierski \cite{bl} proposed an $OSp(1|4n)$--invariant exotic BPS
superparticle in a flat tensorial superspace preserving all but
one supersymmetry of the target superspace, for instance, 3/4 SUSY
in $N=1$, $D=4$ superspace \footnote{BPS states preserving
${{2n-1}\over{2n}}$ supersymmetries (with $n=16$ for $D=10,11$)
have lateron been shown to be building blocks of any BPS state and
conjectured to be hypothetical constituents or `preons' of
M-theory \cite{preons}.}. One of the motivations for Bandos and
Lukierski was a generalization of the Penrose twistor program to
tensorial superspaces and associated superalgebras with tensorial
charges . But it happened that this model turned out to be the
first dynamical realization of the Fronsdal proposal.

Quantum states of the tensorial superparticle was shown
 to form an infinite series of massless higher spin states in
 $D=4$ and first quantized field equations for wave functions
in tensorial superspace have been obtained \cite{exotic}. In
\cite{preit} quantum superparticle dynamics on $OSp(1|4)$ was
assumed to describe higher spin field theory in $N=1$ super
$AdS_4$.

In \cite{exotic} it was shown explicitly how the alternative
Kaluza--Klein compactification produces higher spin fields. It
turns out that in the tensorial superparticle model, in contrast
to the conventional Kaluza--Klein theory, the compactification
occurs in the momentum space and not in the coordinate space.  The
coordinates conjugate to the compactified momenta take discrete
(integer and half integer values) and describe spin degrees of
freedom of the quantized states of the superparticle in
conventional space--time.

In \cite{misha} M. Vasiliev has extensively developed this subject
by having shown that the first--quantized field equations in
tensorial superspace of a bosonic dimension $n(2n+1)$ and of a
fermionic dimension $2nN$ are $OSp(N|4n)$ invariant, and for $n=2$
correspond to so called unfolded higher spin field equations in
$D=4$. It has also been shown \cite{misha1} that the theory
possesses properties of causality and locality.

An alternative derivation that the quantized dynamics of the
superparticle in flat tensorial superspace and on $OSp(1|4)$
reproduces, respectively, the unfolded higher spin field dynamics
in flat $D=4$ superspace and in $N=1$ $AdS_4$ has been given in
\cite{mps}, where the $GL(2n)$ flatness of $OSp(1|2n)$ manifolds
has been observed and used to quantize the $OSp(1|4)$ model.

\section{GL flatness versus conformal flatness}
Before introducing the notion of GL flatness let us remind what
the (super)conformal flatness of supermanifolds is.

A supermanifold is called superconformally flat if its
supervielbeins differ from the flat supervielbeins by a conformal
factor $e^{\rho(x,\theta)}$ (possibly, up to Lorentz rotations).
The vector supervielbein has the following form
\begin{equation}\plabel{cfv}
E^a=e^{\rho(x,\theta)}(dx^a-id\bar\theta\gamma^a\theta)\equiv
e^{\rho(x,\theta)}\Pi^a, \quad a=0,1,\cdots,D-1
\end{equation}
and the spinor supervielbein is
\begin{equation}\plabel{cfs}
E^\alpha=e^{{\rho(x,\theta)}\over 2}(d\theta^\alpha+{i\over
2}\Pi^a\gamma^{\a\b}_aD_\b\rho),
\end{equation}
where
$D_\beta=\partial_\b+i\bar\theta^\d\gamma^a_{\d\b}\partial_a$ is
the flat supercovariant derivative.

An example of the superconformally flat manifolds is $N=1$ super
$AdS_4$ which is a coset superspace ${OSp(1|4)}\over{SO(1,3)}$. A
detailed analysis, along with a criteria for supermanifolds to be
superconformally flat, and a list of such superspaces has been
given in \cite{conflat}.

Recall that the bosonic AdS metric is conformally flat, i.e. can
be presented in the  form $ ds=e^{2\rho(x)}dx^m\,dx^n\eta_{mn}\,.
$

\subsection{GL(2n) flatness of tensorial supermanifolds}
We shall call a tensorial supermanifold GL flat if its
supervielbeins differ from the supervielbeins of flat tensorial
superspace by a GL-group rotation
\begin{eqnarray}\plabel{glf}
&\Omega^{\a\b}=(dx^{\gamma\delta}-{i\over
2}d\theta^\gamma\theta^\delta-{i\over
2}d\theta^\delta\theta^\gamma)\,{\cal G}_{\gamma}^{~\a}(x,\theta)\,{\cal G}_{\delta}^{~\b}(x,\theta)\\
&E^\a=e^{\rho(x,\theta)}({\cal D}\theta^\alpha-\theta^\alpha{\cal
D}\rho)\nonumber,
\end{eqnarray}
where $dx^{\gamma\delta}-{i\over
2}d\theta^\gamma\theta^\delta-{i\over
2}d\theta^\delta\theta^\gamma$ is a flat tensorial space
supervielbein, ${\cal G}_{\gamma}^{~\a}(x,\theta)$ is a general
linear matrix and ${\cal D}$ is a covariant differential.

Apart from the flat space, the only example of GL-flat tensorial
superspaces known to us is the example of supergroup manifolds
$OSp(1|2n)$.
\subsection{$OSp(1|2n)$ supergroup manifolds}
A group element ${\cal O}(x,\theta)$ of $OSp(1|2n)$ is
parametrized by bosonic symmetric $2n\times 2n$ matrices
$x^{\a\b}$ and fermionic variables $\theta^\a$. The Cartan forms
are defined as usual
\begin{equation}\plabel{cf}
{\cal O}^{-1}d{\cal O}=\Omega^{\a\b}(x,\theta)M_{\a\b}+E^\a Q_\a
\end{equation}
and take values in the $OSp(1|2n)$ algebra formed by $Sp(2n)$
generators $M_{\a\b}=M_{\b\a}$ and by their supersymmetric
partners $Q_\a$, such that $\{Q_\a,Q_\b\}=M_{\a\b}$.

The Cartan forms satisfy the Maurer--Cartan equations
\begin{equation}\plabel{mce}
d\Omega^{\a\b}+{\varsigma\over 2}\Omega^{\a\g}\wedge
\Omega_\g^{~\b}=-E^\a\wedge E^\b\,, \quad dE^\a+{\varsigma\over
2}E^\g\wedge\Omega_\g^{~\a}=0,
\end{equation}
where $\varsigma$ is a dimensional parameter which in the case of
$OSp(1|4)$ can be associated with the inverse radius of $N=1$
$AdS_4$, or equivalently with the square root of the cosmological
constant absolute value. When $\varsigma \rightarrow 0$ the
tensorial superspace becomes flat. So $\varsigma$ plays the role
of a contraction parameter. From a physical point of view
$\varsigma$ appeared in the commutation relations of the
$OSp(1|2n)$ superalgebra because we would like to endow the
coordinates $x^{\alpha\beta}$ with a conventional dimension of
length $\varsigma^{-1}$. Then the generators $M_{\a\b}$ have a
dimension of $\varsigma$ and $Q_\a$ that of $\varsigma^{1/2}$.

It turns out that it is possible to choose such a parametrization
of $OSp(1|2n)$ that its Cartan forms become GL-flat \cite{mps} as
in \p{glf}, with
\begin{equation}\plabel{calG}
{\cal G}_\b^{~\a}(x,\theta)=G_\b^{~\a}(x)-{{i\varsigma}\over
8}\left(\Theta_\b-2G_\b^{~\g}(x)\Theta_\g\right)\Theta^\a, \quad
G_\b^{-1\a}(x)=\delta^{~\a}_\b+{\varsigma\over 4}x^{~\a}_\b,
\end{equation}
$\Theta^\a(\theta)$ being defined as the inverse function of
$\theta^\a=\Theta^\b\,G_\b^{-1\a}(x)e^{-\rho(\Theta)}$ and
$e^{\rho(\Theta)}=\sqrt{1 +\frac{i\varsigma}{8}\Theta^\beta
\Theta_\beta}$.

The GL--flatness of bosonic manifolds and supermanifolds seems
novel and is quite interesting also from the mathematical point of
view, and this should be appreciated yet. So far this property has
found a physical application to the quantization of a
superparticle on the group manifold $OSp(1|4)$ (and in general on
$OSp(1|2n)$), and helped to find an explicit solution of the
tensorial field equations and to demonstrate its relation to
higher spin field theory in $AdS_4$. For the sake of simplicity we
will henceforth consider only a non--supersymmetric particle model
in tensorial spaces associated with four-dimensional space-time.
The quantization of this model \cite{exotic,mps} gives rise to a
free higher spin field theory in four-dimensional flat and AdS
spaces \cite{m1}.

\section{Twistor-like particle dynamics in tensorial spaces }
The action proposed in \cite{bl} to describe a particle
propagating in a ten--dimensional tensorial space is
\begin{equation}\plabel{ac}
S=\int\, \Omega^{\a\b}(x(\tau))\l_\a\l_\b,\quad \a=1,2,3,4
\end{equation}
where $\l_\a(\tau)$ is an auxiliary commuting Majorana spinor
variable and $\Omega^{\a\b}(x(\tau))$ is the pullback on the
particle worldline of the tensorial space vielbein.

When $\Omega^{\a\b}(x(\tau))=d\tau\partial_\tau
x^{\a\b}=dx^{\a\b}(\tau)$ we deal with flat tensorial space, and
when $\Omega^{\a\b}(x(\tau))=dx^{\g\d}(\tau)\,{
G}_{\gamma}^{~\a}(x)\,{G}_{\delta}^{~\b}(x)$ with inverse of
$G_{\gamma}^{~\a}(x)$ defined in eq. \p{calG}, the particle
propagates on the group manifold $Sp(4)$.

It is now easy to realize that because of the GL flatness of the
$Sp(4)$ manifold, particle dynamics in flat tensorial space and in
$Sp(4)$ are related to each other by a simple redefinition of
$\l_\a\rightarrow \tilde\l_\a=G_\a^{~\b}(x)\l_\b $ and hence are
classically equivalent
\begin{equation}\plabel{eq}
S=\int\, \Omega^{\a\b}(x(\tau))\l_\a\l_\b =\int\,
dx^{\g\d}(\tau)\,{ G}_{\gamma}^{~\a}(x)\,{
G}_{\delta}^{~\b}(x)\l_\a\l_\b= \int\,
dx^{\a\b}(\tau)\tilde\l_\a\tilde\l_\b\,.
\end{equation}
Without going into details which the reader may find in
\cite{bl,mps}, let us note that the action \p{ac} is invariant
under $Sp(8)$ transformations acting non--linearly on $x^{\a\b}$
and $\l_\a$, i.e. possesses the symmetry which Fronsdal considered
to be an underlying symmetry of higher spin field theory in $D=4$
\cite{fronsdal1}. A group theoretical reason behind the $Sp(8)$
invariance of \p{ac} is that the flat tensorial space and $Sp(4)$
are different realizations of a coset space
${{Sp(8)}\over{GL(4)\times\!\!\!\!\supset K}}$, where $K_{mn}$ are
tensorial analogs of conformal boosts \cite{mps}.

\subsection{Hamiltonian analysis and dynamical properties}
Because of GL flatness and classical equivalence of particle
dynamics in flat tensorial space and in $Sp(4)$ we will first
perform the Hamiltonian analysis and the quantization of the both
cases in the ``flat" basis, which will allow us to understand  the
physical content of the model in the simplest way.

 The particle momenta conjugate to the tensorial coordinates
$x^{\a\b}={1\over 2} x^m\gamma_m^{\a\b}+{1\over 4}
y^{mn}\gamma_{mn}^{\a\b}$ are constrained to be bilinear in
$\l_\a$
\begin{equation}\plabel{mo}
{{\d S }\over{\d x^{\a\b} }}=P_{\a\b}={1\over
2}P_m\gamma^m_{\a\b}+{1\over
4}Z_{mn}\gamma^{mn}_{\a\b}=\l_\a\l_\b.
\end{equation}
As a consequence of \p{mo} the $D=4$ part $P_m$ of the momenta is
expressed via the Cartan--Penrose (twistor) relation and therefore
is light--like in virtue of gamma--matrix Fierz identities in
$D=3,4,6$ and 10
\begin{equation}\plabel{cp}
P_m={1\over 2}\l\gamma_m\l\, \quad \Rightarrow \quad P_mP^m = 0.
\end{equation}
Hence, from the perspective of $D=4$ space--time the particle is
massless.

The constraints which restrict the dynamics of the particle are
\begin{equation}\plabel{con}
D_{\a\b}=P_{\a\b}-\l_\a\l_\b=0, \qquad y^\a=0,
\end{equation}
 where $y^\a$ \footnote{We have called this momentum $y^\a$ to indicate
that this variable is related to one which appears in the Vasiliev
unfolded formulation of higher spin fields
\cite{m1,misha,misha1}.} is the momentum conjugate to $\l_\a$. It
is zero because in \p{ac} $\l_\a$ does not have the kinetic term.

In view of the canonical Poisson brackets
\begin{equation}\plabel{pbr}
[P_{\a\b},x^{\g\d}]={1\over
2}(\delta^\g_\a\delta^\d_\b+\delta^\g_\b\delta^\d_\a)\,, \quad
[y^\a,\lambda_\b]=-\delta^\a_\b\,,
\end{equation}
the constraints obey the following Poisson brackets
\begin{equation}\plabel{pb}
[D_{\a\b},D_{\g\d}]_{PB}=0, \quad [y^\a,y^\b]_{PB}=0, \quad
[y^\a,D_{\b\g}]_{PB}=\delta^\a_\b\l_\g+\delta^\a_\g\l_\b\not = 0.
\end{equation}
 From \p{pb} we conclude that the constraints \p{con} are a mixture
of the first and second class constraints. To quantize the theory
it is easier to work with systems which have only first class
constraints.

Note that the change of variables performed in the $Sp(4)$ action
\p{eq} to pass to the flat basis corresponds to the following
canonical transformation of the Hamiltonian variables which does
not change the canonical Poisson brackets \p{pbr} of the new
variables
\begin{equation}\plabel{ct}
x^{\a\b}=\tilde x^{\a\b}\,,\quad P_{\a\b}=\tilde
P_{\a\b}-{\varsigma\over 8}(\tilde\lambda_\a y_\b+\tilde
\lambda_\b y_\a)\,, \quad \l_\a=G_\a^{-1\b}(x)\tilde\l_\b\,, \quad
y^\a=\tilde y^\b G^{~\a}_{\b}(x)\,.
\end{equation}
Note also that the transformed momentum $\tilde P_{\a\b}$
coincides with the initial one up to the terms which are
proportional to the constraint $y_\a=0$ and hence are weekly equal
to zero.

To pass to a system with only first class constraints which is
physically equivalent to the original one we should make a
conversion of the constraints \p{con} into the first class in such
a way that the number of physical degrees of freedom remains the
same. In our case the conversion procedure is very simple. One
just promotes $y^\a$ to an unconstrained dynamical variable. Then
the remaining constraints $D_{\a\b}$ are of the first class and
generate local worldvolume symmetries of the action \p{ac}, while
the condition $y^\a=0$, when imposed, is regarded as gauge fixing
of a part of these local symmetries. The conversion procedure
described above is equivalent to adding to the action \p{ac} the
first--order kinetic term $\int d\l_\a(\tau)\,y^\a(\tau)$ for
$\l_\a$ (see e.g. \cite{misha}).
\subsection{Quantization and field equations in flat tensorial space}
Upon the conversion the quantization of particle dynamics is
straightforward. One should promote the dynamical variables to
operators, to replace the Poisson brackets with commutators and to
impose the first class constraints on the particle wave function.

One can consider the particle wave function in different (momentum
and/or coordinate) representations related to each other by the
Fourier transform.

For instance, in the representation considered in \cite{bl}, which
we shall call the $\l$--represen\-tation, the wave function
$\Phi(x,\l)$ is assumed to depend on $x^{\a\b}$ and $\l_\a$, while
$P_{\a\b}={\partial\over{i\partial x^{\a\b}}}$ and
$y^\a=i{\partial\over{\partial\l_\a}}$ are realized as
differential operators. The wave function satisfies the equation,
which is the quantum counterpart of the first class constraints
\cite{exotic}
\begin{equation}\plabel{l}
D_{\a\b}\Phi(x,\l)=\left({\partial\over{\partial
x^{\a\b}}}-i\l_\a\l_\b\right)\Phi(x,\l)=0\,.
\end{equation}
The general solution of \p{l} is very simple
\begin{equation}\plabel{soll}
\Phi(x,\l)=e^{ix^{\a\b}\l_\a\l_\b}\varphi(\l),
\end{equation}
where $\varphi(\l)$ is a generic function of $\l_\a$.

We can now make the Fourier transform of \p{soll} to another
representation to be called $y$--representation
\begin{equation}\plabel{yr}
C(x,y)=\int\,d^4\l\,e^{-iy^\a\l_\a}\Phi(x,\l)=\int\,d^4\l\,e^{-iy^\a\l_\a+
i x^{\a\b}\l_\a\l_\b}\varphi(\l).
\end{equation}
The wave function $C(x,y)$ satisfies the Fourier transformed eq.
\p{l}
\begin{equation}\plabel{y}
\left({\partial\over{\partial
x^{\a\b}}}+i{\partial^2\over{\partial y^\a\partial
y^\b}}\right)C(x,y)=0.
\end{equation}
This equation has been analyzed in \cite{misha} for the wave
functions which are polynomials in $y^\a$
\begin{equation}\plabel{pol}
C(x,y)=\sum^\infty_{n=0}C_{\a_1\cdots\a_n}(x)y^{\a_1}\cdots
y^{\a_n}=b(x)+f_\a(x)y^\a+\cdots\,.
\end{equation}
Substituting \p{pol} into \p{y} one finds that the scalar field
$b(x)$ and the spinor field $f_\a(x)$ satisfy the following
equations
\begin{equation}\plabel{bf}
(\partial_{\a\b}\partial_{\g\d}-\partial_{\a\g}\partial_{\b\d})b(x)=0,
\quad \partial_{\a\b} f_\g(x)-\partial_{\a\g} f_\b(x)=0,
\end{equation}
so these fields are dynamical, while all higher components in the
expansion \p{pol} are expressed in terms of (higher) derivatives
of $b(x)$ and $f_\a(x)$ and, hence, are auxiliary fields.

Thus the quantum dynamics of the particle in tensorial spaces is
described by only two dynamical fields, which can be obtained from
the wave function in the $\lambda$--representation by integrating
the latter over $\l_\a$ as follows
\begin{equation}\plabel{bfl}
b(x)=\int\,d^4\l\,\Phi(x,\l), \qquad
f_\a(x)=-i\int\,d^4\l\,\l_\a\,\Phi(x,\l)\,.
\end{equation}
In virtue of the GL flatness of $Sp(4)$ the consideration above is
applicable both to flat space and to $Sp(4)$, though in the latter
case, because of the field redefinition \p{eq}, $x^{\a\b}$ and
$\tilde\l_\a$ transform under $Sp(4)$ in a highly non--linear way.
So, $Sp(4)$ symmetry is not manifest. We are in a similar
situation to that of a `free-fermion' model with a non--linearly
realized $SU(n|1)$ symmetry considered in \cite{luka}. To restore
manifest $Sp(4)$ invariance we should return to original variables
at the expense of the loss of the `free' character of dynamics.

\subsection{Particle dynamics on Sp(4)}
The Hamiltonian constraints which follow from \p{eq} (without
doing the GL rotation) have the form
\begin{equation}\plabel{spc}
y^\a=0\,,\quad
D_{\a\b}=G_\a^{-1\g}(x)G_\b^{-1\d}(x)
P_{\g\d}-\l_\a\l_\b=\nabla_{\a\b}-\l_\a\l_\b=0
\,,
\end{equation}
where $\nabla_{\a\b}=G_\a^{-1\g}(x)G_\b^{-1\d}(x)P_{\g\d}$
generate the $Sp(4)$ algebra
$[\nabla_{\a\b},\nabla_{\g\d}]_{PB}=-{\varsigma \over 2}
C_{\a\{\g}\nabla_{\d\}\b}-{ \varsigma \over 2}
C_{\b\{\g}\nabla_{\d\}\a}$, and $C_{\a\b}$ is a simplectic metric.
(Remember that $G_\a^{-1\g}(x)=\delta^\g_\a+{\varsigma\over
4}x^{~\g}_\a$ and $G_\a^{-1\g}(x)G_\b^{-1\d}(x)$ is inverse of the
Cartan form matrix \p{eq}, i.e. the inverse vielbein of the group
manifold $Sp(4)$).

Because of the non--commutativity of $\nabla_{\a\b}$ the
constraints $D_{\a\b}$ do not commute even in the weak Dirac
sense, i.e. $[D_{\a\b},D_{\g\d}]_{PB} \not =0$. This is in
contrast to what we had in the flat case \p{pb}. However the weak
commutativity can be restored if we modify $D_{\a\b}$ by adding to
them terms linear and quadratic in the constraint $y^\a=0$ as
follows
\begin{equation}\plabel{calD}
{\cal D}_{\a\b}=\nabla_{\a\b}-(\l_\a+{\varsigma\over 8}
y_\a)(\l_\b+{\varsigma\over 8} y_\b)=\nabla_{\a\b}-Y_\a Y_\b=0\,.
\end{equation}
The constraints \p{calD} can be obtained from the flat constraints
\p{con} performing the canonical transformations \p{ct} and adding
to \p{con} appropriate terms linear and quadratic in the
constraint $y_\a$, which can always be done. At the classical
level the addition of these terms is just another choice of
constraints, however at the quantum level this changes background
geometry (in our case from flat tensorial space to $Sp(4)$). It
occurs in the following way.

Since $Y_\a\equiv \l_\a+{\varsigma\over 8} y_\a$  do not commute
and $[Y_\a,Y_\b]_{PB}={\varsigma\over 4} C_{\a\b}$, the
constraints ${\cal D}_{\a\b}$, like $\nabla_{\a\b}$,  generate the
$Sp(4)$ algebra $[{\cal D}_{\a\b},{\cal
D}_{\g\d}]_{PB}=-{\varsigma \over 2} C_{\a\{\g}{\cal
D}_{\d\}\b}-{\varsigma \over 2} C_{\b\{\g}{\cal D}_{\d\}\a}= 0$
and hence weakly commute. Thus, the constraints ${\cal D}_{\a\b}$
reflect the $Sp(4)$ structure of the tensorial space where the
particle propagates.

As in the flat case \p{pb} $y^\a$ do not commute with ${\cal
D}_{\a\b}$ and the whole system of the constraints is again a
mixture of the first and second class ones. As before we convert
it into the first class by regarding $y^\a$ to be unconstrained.

Now the quantization of the system is performed as in the previous
Subsection. The first--class constraints become operators which
annihilate physical states of the particle on $Sp(4)$. In the
$\l$--representation the first quantized wave function satisfies
the equation
\begin{equation}\plabel{lsp}
{\cal D}_{\a\b}\Phi(x,\l)=\left[\nabla_{\a\b}-{i\over 2}(Y_\a
Y_\b+Y_\b Y_\a)\right]\Phi(x,\l)=0\,, \quad Y_\a\equiv
\l_\a+{i\varsigma\over 8} {\partial\over {\partial\l^\a}}\,,
\end{equation}
and in the $y$--representation
\begin{equation}\plabel{ysp}
{\cal D}_{\a\b}C(x,y)=\left[\nabla_{\a\b}-{i\over 2}(Y_\a
Y_\b+Y_\b Y_\a)\right]C(x,y)=0\,\quad Y_\a\equiv {i}
{\partial\over {\partial y^\a}}+{\varsigma\over 8}y_\a\,,
\end{equation}
where $\nabla_{\a\b}$ is a covariant derivative on $Sp(4)$.

We see that on $Sp(4)$ the two representations are completely
equivalent, or dual, to each other with respect to the exchange of
$\l_\a$ and ${\varsigma  \over 8}y_\a$, which is reflected in the
form of the general solutions of these equations.

Symmetries and solutions of eq. \p{ysp} have been studied in
\cite{dv}. The GL--flat realization of the $Sp(4)$ Cartan forms
\p{eq} and of the covariant derivatives \p{spc} allows us to find
the general solutions of \p{lsp} and \p{ysp} in a very simple form
akin to that of the flat case \p{soll}, \p{yr}
\begin{equation}\plabel{ls}
\Phi(x^{\a\b},\l)  =\int\, d^4y\, \sqrt{\det G^{-1}(x)}\,
e^{{i}x^{\a\b}(\l_\a+ {\varsigma\over 8}y_\a
)(\l_\b+{\varsigma\over 8}y_\b) +i \l_\a y^\a}\,\varphi(y)\,,
\end{equation}
\begin{equation}\plabel{ys}
C(x^{\a\b},y)=\int\, d^4\l\, \sqrt{\det G^{-1}(x)}\,
e^{{i}x^{\a\b}(\l_\a+ {\varsigma\over 8}y_\a
)(\l_\b+{\varsigma\over 8}y_\b) -i \l_\a y^\a}\,\varphi(\l)\,.
\end{equation}
To find \p{ls} and \p{ys} we have used that
\begin{eqnarray}\plabel{dG}
& G^{-1\b}_\a(x)=\delta^\b_\a+{\varsigma\over 4}
x^{~\b}_\a\,,\quad \nabla_{\a\b}\,\det
G^{-1}=G^{-1\a'}_\a\,G^{-1\b'}_\b {\partial \det
G^{-1}\over{\partial x^{\a'\b'}}}={\varsigma^2\over 16}\,
x_{\a\b}\,\det
G^{-1}\,,\\
 & \nabla_{\a\b}G^{\g\d}={\varsigma\over
4}(\d^\g_{\{\a} + 2G^\g_{~\{\a})\d^\d_{\b\}}\,. \nonumber 
\end{eqnarray}
For completeness let us also present the explicit form of the
$\det{G^{-1}}$:
\begin{equation}\plabel{det}
\det{G^{-1}}=1-{1\over 2}\,\left({\varsigma\over 4}\right)^2 \,
\,x_\a^{~\b}x^{~\a}_\b+{1\over 8}\,\left({\varsigma\over
4}\right)^4 \,(x_\a^{~\b}x^{~\a}_\b)^2-{1\over 4}\,
\left({\varsigma\over 4}\right)^4 \,x_\a^{~\b}x^{~\g}_\b
x^{~\d}_\g x^{~\a}_\d\,.
\end{equation}

 One can wonder what is the $Sp(4)$ analog of the equations
\p{bf} of the dynamical fields $b(x)$ and $f_\a(x)$ entering the
polynomial wave function \p{pol}. Upon some algebra we arrive at
the following system of equations
\begin{equation}\plabel{bsp}
\nabla_{\a[\b}\nabla_{\g]\d}b(x)={\varsigma\over
16}\left(C_{\a[\b}\nabla_{\g]\d}- C_{\d[\g}\nabla_{\b]\a} +
2C_{\b\g}\nabla_{\a\d}\right)b(x)+{\varsigma^2\over
64}\left(2C_{\a\d}C_{\b\g}-C_{\a[\b}C_{\g]\d}\right)b(x),
\end{equation}
\begin{equation}\plabel{fsp}
\nabla_{\a[\b}f_{\g]}(x)=-{\varsigma\over
4}\left(C_{\a[\g}f_{\b]}(x)+2C_{\b\g}f_\a(x)\right)\,.
\end{equation}
These $Sp(4)$ equations can be regarded as tensorial counterparts
of the equations of motion of a massless scalar and spinor field
in $AdS_4$.

We should stress that the consideration and the formulas presented
in Section 4 with the example of particle dynamics on $Sp(4)$
remain the same for a generic case of the group manifold $Sp(2n)$.
In particular, in $Sp(2n)$ the wave equations \p{lsp}, \p{ysp},
\p{bsp} and \p{fsp} and their solutions have the same form as
above. When $n=1$ the group manifold $Sp(2)\sim SO(1,2)$ is
isomorphic to $AdS_3={{SO(2,2)}\over{SO(1,2)}}$, and the equations
\p{bsp} and \p{fsp} reduce to the well--known equations of motion
of a massless scalar and spinor field in $AdS_3$.

\section{Higher spin fields in ordinary $D=4$ space--time} Let us
now demonstrate how higher spin fields and their unfolded field
equations emerge in the ordinary four--dimensional subspace of the
tensorial space. For this we should just formally rewrite the wave
function \p{yr} in a different way, namely
\begin{equation}\plabel{yr4}
C(x^{\a\b},y^{\a})=\int\,d^4\l\,e^{-iy^\a\l_\a+ix^{\a\b}\l_\a\l_\b}\varphi(\l)=
\int\,d^4\l\,e^{-i\tilde y^\a\l_\a+{i\over 2}
x^m\g_m^{\a\b}\l_\a\l_\b}\varphi(\l)=C(x^m,\tilde y^\a),
\end{equation}
where we have  used the coordinate decomposition \p{x} and have
hidden the six tensorial coordinates $y^{mn}$ into  redefined
$\tilde y^\a=y^\a-{1\over 4}y^{mn}\g_{mn}^{\a\b}\l_\b$
\footnote{Note that $x^{\a\b}$ can be completely absorbed by a
redefined $y^\a$, then we recover a twistor--like transform of the
tensorial space \cite{bl,exotic}.}. In other words, what we have
actually done is we have taken $C(x^{\a\b},y^{\a})$ at $y^{mn}=0$
.

By construction the wave function $C(x^m,\tilde y^\a)$ satisfies
the field equations in $D=4$ space--time
\begin{equation}\plabel{4df}
\left({\partial\over{\partial x^m}}+{i\over
2}\gamma_m^{\a\b}{\partial^2\over{\partial \tilde y^\a\partial
\tilde y^\b }}\right)C(x^m,\tilde y^\a)=0\,.
\end{equation}
These are unfolded field equations of Vasiliev \cite{m1,misha}
which produce the flat $D=4$ equations of motion of the field
strengths of the higher spin ($s=n/2$) fields. The field strengths
are the components $C_{\a_1\cdots\a_n}(x^m)$ of the polynomial
expansion of $C(x^m,\tilde y^\a)$ (called the generating function)
\begin{equation}\plabel{gf}
C(x^m,\tilde
y^\a)=\sum^\infty_{n=0}C_{\a_1\cdots\a_n}(x^m)\,\tilde
y^{\a_1}\cdots \tilde y^{\a_n}\,.
\end{equation}
To obtain the generating function of higher spin fields in $AdS_4$
we should just take the solution \p{ys} at $y^{mn}=0$, which then
takes the form
\begin{equation}\plabel{yads}
C(x^m,y^a)=\int\, d^4\l\, (1-{\varsigma^2\over{8^2}}\,x^mx_m)\,
e^{{i\over 2 }x^m\g_m^{\a\b}(\l_\a+ {\varsigma\over 8}y_\a
)(\l_\b+{\varsigma\over 8}y_\b) -i \l_\a y^\a}\,\varphi(\l)\,,
\end{equation}
where now $G^{-1\b}_\a(x^m)=\delta_\a^\b+{\varsigma\over
8}x^m\gamma_{m\a}^{~~\b}$ and $\det
G^{-1}(x^m)=(1-{\varsigma^2\over{64}}\,x^mx_m)^2$.

To get the unfolded $AdS_4$ equations satisfied by \p{yads} we
should multiply equations \p{lsp} or \p{ysp} by ${1\over 2}
G^{~\a}_{\d}\,\gamma_m^{\d\sigma}G_{\sigma}^{~\b}$ and then take
$G^{~\a}_{\d}(x^m)$ at $y^{mn}=0$, we thus get
\begin{equation}\plabel{adssp}
\left[{\partial\over{\partial
x^m}}+i\Omega^{\a\b}_m(x)Y_\a\,Y_\b\right]C(x^m,y^\a)=0,
\end{equation}
where
\begin{equation}\plabel{Omega}
\Omega^{\a\b}(x^m)=dx^m\,\Omega_m^{\a\b}(x^m)={1\over
2}dx^m\gamma_{m}^{\d\sigma}G^{~\a}_{\d}(x)\,G_{\sigma}^{~\b}(x) =
{1\over 4}dx^m\,\omega_m^{ab}(x)\,\gamma_{ab}^{\a\b}+{1\over 2}
dx^m\,e_m^a(x)\gamma_a^{\a\b}
\end{equation}
is the generalized $AdS_4$ connection satisfying the zero
curvature condition $d\Omega+{\varsigma\over
2}\,\Omega\wedge\Omega=0$. It is composed of the $AdS_4$ spin
connection $\omega_m^{ab}(x)$ and the $AdS_4$ vielbein $e_m^a(x)$,
and hence takes values in $Sp(4)$.

$C(x^m,y^a)$ of \p{yads} is the generating function of the field
strengths of higher spin fields in $AdS_4$. Its form is different
from that considered in \cite{BV,mps} because the latter was
written in the $AdS_4$ parametrization in which the $AdS$ metric
is {\it conformally flat}, while eq. \p{yads} is in the {\it
GL--flat basis} of the $AdS$ isometry group $Sp(4)$.

\section{Conclusion and discussion} Having based upon results of
\cite{exotic,preit,misha,mps} we have demonstrated how free higher
spin field theory in $D=4$ flat space-time and in $AdS_4$ emerges
upon the quantization of a simple particle model \cite{bl},
respectively, in flat tensorial space and on the group manifold
$Sp(4)$ generating isometries of $AdS_4$.

To analyze the model we have used the property of these tensorial
spaces to be $GL(4)$-flat, which is a tensorial analog of the
conformal flatness of the Minkowski and $AdS$ spaces.

As a generalization and development of these results, higher
dimensional and supersymmetric $OSp(N|2n)$ extensions of tensorial
particle dynamics and corresponding first--quantized higher--spin
field theories have been studied in \cite{bl}--\cite{mps},
\cite{dv,fedor,m2}.

The dynamics of extended relativistic objects in tensorial
superspaces has been analyzed as well. For instance tensionless
("null") superbranes in tensorial spaces have been considered in
\cite{uzh,i,uzhe} and fully fledged superstrings in
\cite{curt,i1}. Group--theoretical aspects of brane dynamics
involving tensorial charges have been discussed in \cite{mm}.

In conclusion let us note that in spite of a progress in
understanding the subject considered above, as in most of the
theoretical fields, many questions are still to be answered, for
instance
\begin{itemize}
\item
Do there exist other GL--flat (super)manifolds in addition to
$OSp(1|2n)$, \linebreak e.g. $OSp(N|2n)$ with $N>1$?
\item
Do field equations \p{bf}, \p{bsp} and \p{fsp} in tensorial spaces
admit a Lagrangian interpretation, i.e. whether they can follow
from an action principle? Note that the equations are rather
simple but highly degenerate.
\item
 Whether particle and field dynamics in curved tensorial
spaces may be helpful in solving the Interaction Problem of Higher
Spin Field Theory? A step in this direction was made in \cite{r}
where cubic interactions of fields in tensorial spaces were
analyzed.
\item
Is there any relation of field theory in tensorial spaces to
higher spin field theory produced by strings
\cite{fs,bpt,bonelli}?

\end{itemize}
{\bf Acknowledgements.} The authors are grateful to I. Bandos, X.
Bekaert, J. Lukierski and M. Vasiliev for interest to this work
and useful discussions. This work was partially supported by the
Grants 1010073 and 7010073 from FONDECYT (Chile) (M.P. and D.S.)
and by DICYT (USACH) (M.P.), by the Grant N 383 of the Ukrainian
State Fund for Fundamental Research (D.S.), by the INTAS Research
Project N 2000-254 and by the European Community's Human Potential
Programme under contract HPRN-CT-2000-00131 Quantum Spacetime
(D.S. and M.T.).

\end{document}